\documentstyle[epsf,12pt,aaspp]{article}
\tighten

\def\bohringer	{B\"{o}hringer}
\def\muller	{M\"{u}ller}
\def\am		{$^\prime$}

\def\kmsmpc	{~km$\;$s$^{-1}\,$Mpc$^{-1}$}
\def\msun	{~$M_{\odot}$}
\def\kms	{~km$\;$s$^{-1}$}

\def\gax	{{_{\textstyle >}\atop^{\textstyle \sim}}}
\def\ergs	{~erg$\;$s$^{-1}$}
\def\cmcube	{~cm$^{-3}$}

\begin{document}

\title{Abell 2163: temperature, mass, and hydrostatic equilibrium}

\vspace{1cm}

\author{M.~Markevitch\altaffilmark{1,4}, R.~Mushotzky\altaffilmark{2},
H.~Inoue\altaffilmark{1}, K.~Yamashita\altaffilmark{3},
A.~Furuzawa\altaffilmark{3}, Y.~Tawara\altaffilmark{3}}

\altaffiltext{1}{ISAS, 3-1-1 Yoshinodai, Sagamihara, Kanagawa~229, Japan.
maxim, inoue @astro.isas.ac.jp}

\altaffiltext{2}{NASA/GSFC, Greenbelt, MD 20771, USA. mushotzky@gsfc.nasa.gov}

\altaffiltext{3}{Department of Physics, Nagoya University, Furo, Chikusa,
Nagoya 461-01, Japan. yamasita, furuzawa, tawara @phys.nagoya-u.ac.jp}

\altaffiltext{4}{Permanent address: IKI, Profsoyuznaya 84/32, Moscow 117810,
Russia. maxim@hea.iki.rssi.ru}

\vspace{2cm}\centerline{Accepted for {\em The Astrophysical Journal}}

\clearpage

\begin{abstract}

Using ASCA data, we have measured the electron temperature in Abell 2163 out
to $1.5\,h^{-1}$~Mpc (3/4 of the virial radius, or $10a_x$, where $a_x$ is
the X-ray core-radius) from the center, in three radial bins. The obtained
temperatures are $12.2^{+1.9}_{-1.2}$~keV, $11.5^{+2.7}_{-2.9}$~keV and
$3.8^{+1.1}_{-0.9}$~keV in the 0--3$\,a_x$ (0--3.5\am), 3--6$\,a_x$ and
6--13$\,a_x$ spherical shells, respectively\footnote{Errors are 90\%
throughout the paper unless otherwise stated, and $h\equiv
H_0/100$\kmsmpc.}. Formally applying the hydrostatic equilibrium and
spherical symmetry assumptions and using these data together with the Ginga
spectral and the Rosat imaging data, we were able to severely limit the
possible binding mass distribution of the generic form
$\rho=\rho_0\,(1+r^2/a_b^2)^{-n/2}$. All the allowed binding mass profiles
are steeper than the gas density profiles and mass profiles with the same
slope as gas are excluded at a greater than 99\% confidence.  The total mass
inside $0.5\,h^{-1}$~Mpc is $4.3\pm0.5\,\times10^{14}\,h^{-1}$\msun, of
which $0.074\,h^{-3/2}$ is gas while inside $1.5\,h^{-1}$~Mpc the mass is
$1.07\pm0.13\,\times10^{15}\,h^{-1}$\msun.

The strongest constraint on the mass profile is the observed quick drop of
the temperature at large radii, which can only marginally be reconciled with
the Rosat detection of gas at an even greater radius. We note that in the
outer part of this cluster, which is likely to be a recent merger, the
timescale for reaching electron-ion temperature equality via collisions is
comparable to the merger timescale, so the measured electron temperature may
give an underestimate of the gas pressure there. Otherwise, if our low value
is indeed representative of the gas temperature in the outer shell, the
cluster atmosphere should be convectionally unstable and gas turbulence
should exist. Bulk motions of the gas are also expected during the merger.
Their existense would increase the total gas pressure above that indicated
by the observed temperature. Thus, failure of the model in which dark matter
and gas have the same distribution at the radii of interest, which is
favored by hydrodynamic simulations, may be due to the neglect of these
phenomena, leading to an underestimate of the total density and overestimate
of the baryonic fraction at large radii. The mass estimate at the smaller
radius, where there is no evidence of departing from equilibrium, is likely
to be correct.

Our measured electron temperatures, combined with the previously reported
Sunyaev-Zel'dovich decrement toward this cluster and the Rosat gas density
profile, under the assumption of spherical symmetry, are consistent with a
Hubble constant between 42--110\kmsmpc\ (68\% interval), where the
uncertainty is dominated by that of the available SZ measurement.

\end{abstract}


\clearpage

\section{INTRODUCTION}

After the realization that extended X-ray emission from clusters is due to
hot optically thin plasma (e.g., Mitchell et al.\ 1976; Mushotzky et al.\
1978), it was suggested that measuring the spatial distribution and the
temperature of this emission would allow a relatively robust determination
of the cluster binding mass distribution (e.g., Fabricant, Rybicki \&
Gorenstein 1984), provided that the gas is in hydrostatic equilibrium in the
cluster gravitational well. This method requires a spatially-resolved
temperature measurement, which was impossible until recently. Thus previous
estimates have usually assumed that the cluster gas is isothermal, outside
the sometimes cooler central regions (for the most recent application see
e.g., Elbaz, Arnaud \& \bohringer\ 1995, hereafter EAB, who analyzed A2163).
However, Ginga scanning of the Coma cluster (Hughes 1991), and recent Rosat
spatially resolved temperature measurements (e.g., in A2256, Briel \& Henry
1994) indicate that the cluster temperature structure may be complex,
although approximate isothermality is indeed observed inside the central
several core-radii. Below, we add to these data with an ASCA detection of
the temperature decline in the outer part of Abell 2163, whose temperature
profile appears to resemble that of Coma, the only other cluster at the
moment in which the gas temperature is measured out to a comparable radius.
Moreover, our data imply that the gas cannot be in hydrostatic equilibrium
at the observed temperatures in the outer part of this cluster, which
probably underwent a recent merger (Soucail et al.\ 1995; EAB). Earlier,
using the gravitational lensing and X-ray data, Loeb and Mao (1994)
demonstrated that in another, apparently relaxed cluster, A2218, gas in the
central part cannot be in the pressure equilibrium at the measured
temperature, and a similar problem was noted by Daines et al.\ (1994) in
A1689. These findings may indicate that more physics should be taken into
account before using X-ray measurements to estimate cluster masses.

In this paper, we continue to analyze the August 1993 ASCA observation of
A2163, the hottest and most luminous of known clusters (mean
$T_e=$12--15~keV; EAB, our estimate; $\L_x(2-10~{\rm keV})=6 \times
10^{45}$\ergs, Arnaud et al.\ 1992), situated at $z=0.201$ with the galaxy
radial velocity dispersion of 1680\kms\ (Soucail et al.\ 1995; Arnaud et
al.\ 1994). It possesses the most luminous and extended radio halo of those
detected so far (Herbig \& Birkinshaw 1995). Some preliminary ASCA results,
which include the detection of the $H$-like iron line in the cluster
spectrum and the gas temperature variations in the cluster central part, are
presented in Markevitch et al.\ (1994; hereafter Paper I). Details of the
data reduction relevant to the results reported here, and the obtained
temperature distribution are presented below in Section 2 and in Appendix.
In Section 3, these new temperatures, the Rosat density profile and the
Sunyaev-Zel'dovich measurement by Wilbanks et al.\ (1994) are used to
estimate the Hubble constant. In Section 4, an attempt is made to constrain
the cluster binding mass distribution, formally applying the hydrostatic
equilibrium assumption. In Section 5, we point to the possibilities of the
deviation from the equilibrium.

\section{MEASURING THE TEMPERATURES}

Details of the observation used here are given in Paper I. Our aim now is to
obtain the large-scale temperature distribution in Abell 2163. To do this,
we used data from both the GIS and SIS, restricting our analysis to energies
above 2.5~keV to avoid uncertainty of the excessive absorption toward this
cluster (EAB), and because of the still uncertain instrument calibration at
lower energies. We collected spectra in 3 rings of radii
0--3.5\am--7.5\am--11.5\am\ centered on the cluster brightness peak, in 5
energy bins, 2.5--3--4--5--7--11~keV. They were simultaneously fit by a
model consisting of a spherically-symmetric $\beta$-model gas density
distribution (Jones \& Forman 1984) with its parameters fixed at their Rosat
values ($a_x=1.2'\pm0.075'$ and $\beta=0.62^{+0.015}_{-0.02}$; EAB), with
the constant temperatures in the corresponding spherical shells of radii
0--2.9--6.3--13$a_x$ being free parameters. We chose 13$a_x$ as an outer
radius in the model (which is the maximum radius at which the emission was
actually detected at the 90\% confidence by Rosat; EAB).\ The exact value of
this radius does not significantly affect the results. The outer radius of
the third image ring is chosen smaller than that to reduce the background
contribution and to exclude the uncalibrated detector areas. The Rosat
density profile, derived from the PSPC data (whose angular resolution is
better than that of ASCA) in the assumption of isothermality, is adequate
for this hot cluster even in the presence of significant temperature
variations, since, as the plasma temperature changes from 3 to 15~keV, its
emissivity in the Rosat band changes by only 20\%. We also show below
(Section 2.2) that this density profile is consistent with the ASCA image. A
rudimentary correction of the Rosat emission measure for non-isothermality
has nevertheless been made in each shell (which for the detected temperature
decline roughly corresponds to an increase of $\beta$ to 0.64), which had no
significant effect on the results. Each layer was projected to the sky,
multiplied by the telescope efficiency, convolved with the mirror PSF
dependent on the energy and the focal plane position, and its contribution
to each image ring was calculated for each energy, for the 2 GIS detectors
and 4 SIS chips. The Raymond \& Smith (1977, 1992 version) model has been
used, fixing the redshift at 0.2 (Soucail et al.), the abundance at 0.4
(Paper I) and the absorption at log~$N_H=21.22$ (EAB). Details of the
procedure are given in Appendix. Fig.~1 shows the data values for six
detectors and three image rings, along with the best-fit model. Fig.~2 shows
the resulting best-fit temperatures, $12.2^{+1.9}_{-1.2}$~keV,
$11.5^{+2.7}_{-2.9}$~keV and $3.8^{+1.1}_{-0.9}$~keV in the three spherical
regions, respectively (the errors are 90
parameter). The fit is good, with $\chi^2=69/90-8$~d.o.f.\ (8 parameters
include 3 temperatures, a single normalization for both GISs and 4 separate
normalization for the SIS chips, see below). The confidence intervals given
above are obtained by Monte-Carlo simulation (in 90\% of the trials, the
best-fit value of a temperature was within this interval) and incorporate
errors of the background and the PSF model used. These two most serious
sources of the potential systematic error are detailed below, before
discussing the physical implications of this measurement.

\subsection{Background subtraction}

For the temperature decline in the outer region to be an artifact, the GIS
background has to be about 2 times lower than we assumed. The day-to-day
$1\sigma$ variation of the GIS background rate integrated over the detector,
for the same geomagnetic rigidity value, is 20--30\% (Kubo 1994). We ensured
a better accuracy (about 6\%) by first reconstructing the background from a
set of the blank field observations, normalized according to their exposures
and the distribution of the geomagnetic rigidity during the observation, and
then fitting a normalization of this model background image to the outer
parts of the GIS images of our particular observation. The cluster emission
scattered to the image regions used for the background fit was removed using
a beta-model cluster brightness convolved with the mirror PSF. Obtained
normalization coefficients for both GISs were consistent with 1 (as was
expected for our observation date from the monitoring by Kubo 1994). A
similar method is not applicable to the SIS with its smaller field of view,
and we assumed an error of 20\% on the normalization of the SIS background
calculated from the blank fields, analogous to that of GIS.

\subsection{Mirror PSF}

The mirror PSF has a half-power diameter of about 3\am\ and wide wings whose
shape depends on energy and the focal plane position (Serlemitsos et al.\
1995; Takahashi et al.\ 1995). Because of the wide PSF, the flux in the
outer ring is in fact dominated by that scattered from the inner cluster
part, and the fraction of the scattered flux is energy-dependent. For
example, for our gas density profile and a constant temperature, fractions
of the flux from the inner, middle and outer three-dimensional model shells
are (0.80, 0.18, 0.02) in the inner image ring, (0.41, 0.43, 0.16) in the
second ring, and (0.28, 0.24, 0.48) in the outer ring for $E=2.5$~keV, while
for $E=10$~keV, the respective fractions are (0.79, 0.18, 0.03), (0.43,
0.41, 0.16), and (0.40, 0.24, 0.36). The inward contributions are mainly due
to projection effects, and the outward contributions are because of the PSF
scattering. These numbers illustrate the degree of correlation between the
measured temperatures, and show that even though the image rings are much
larger than the PSF half-power diameter, flux from the inner cluster regions
can be extremely important in the outer rings, while the inward
contributions are small. One also notices that the sign of the energy
dependence of these fractions is such that, if these effects are not
accounted for, the outer part of an isothermal cluster would look hotter.
These effects are largest for the hotter systems with small angular scale
lengths such as Abell 2163.

To take this scattering into account, we have used a two-dimensional model
of the mirror PSF with 1\am\ resolution, obtained by interpolation between a
set of high quality narrow-band GIS images of Cyg~X-1, a bright point source
with a hard spectrum, observed at 11 focal plane positions (Takahashi et
al.\ 1995; Ikebe 1995). This PSF model ignores the differences between the
four instruments and irregularities of each mirror, and cannot be used at
the energies below 2~keV. The model reproduces an actual point source radial
brightness profile reasonably well, and a $1\sigma$ systematic error of 5\%
on the PSF integral over concentric rings of a several-arcminute width has
been assumed. We have added this error to contributions of the inner model
shells to the outer image regions during the Monte-Carlo calculation of the
confidence intervals (see Appendix), conservatively assuming that the error
is the same (100\% correlated) in all energy bands. However, for the SIS, it
may be an underestimate of the error, because neither of the image regions
is contained within one chip, while the PSF uncertainty for integration
regions other than concentric rings are not well understood and are larger,
because of the ignored mirror asymmetries. For SIS, we used the Cyg~X-1
images which have been smoothed to compensate for the relatively
insignificant but energy-dependent intrinsic GIS resolution, and the SIS
cluster images were also smoothed to the same final resolution before
collecting the ring spectra.

An analog of the PSF systematic error was added to the data errors when
calculating the reported best-fit $\chi^2$ values (see Appendix), although
$\chi^2$ is reasonable even without this correction. To confirm that our PSF
model is adequate, we have attempted to free the parameters of the density
profile, $a_x$ and $\beta$, and fit them together with the temperatures in
three model shells to the GIS data in the same rings as above (excluding SIS
for simplicity). The resulting two-parameter confidence contours in the
$a_x-\beta$ plane, obtained by fitting the three temperatures for each pair
of $a_x$ and $\beta$, are presented in Fig.~3, and are perfectly consistent
with the Rosat isothermal fit of the brightness profile. A direct comparison
in the Rosat energy band cannot be performed at the moment since the mirror
PSF below 2~keV is unknown. We conclude that our procedure is adequate and
we are justified to fix $a_x$ and $\beta$ at their EAB's values.

\subsection{SIS/GIS/Ginga consistency}

We analyzed the SIS and GIS data separately, and obtained results which are
in good agreement. Using the GIS only, the temperatures in the three regions
are $11.6^{+2.2}_{-1.2}$~keV, $14.1_{-3.9}^{+4.3}$~keV and
$3.9^{+1.6}_{-0.9}$~keV, with $\chi^2=24.5/30-4$~d.o.f. Using two chips in
each SIS detector (which subtend a $11'\times22'$ rectangle in the sky and
thus only a part of the outer ring is covered), we obtained
$12.6_{-2.3}^{+3.5}$~keV, $8.7_{-2.9}^{+4.1}$~keV and
$3.4_{-1.3}^{+2.4}$~keV, with $\chi^2=39.4/60-7$~d.o.f.\ (all errors are 90%
confidence uncertainties). The cluster is significantly asymmetric; in order
to accommodate for this effect we allowed the relative normalizations of the
four chips to be free parameters. We chose to use only GIS data for the
Hubble constant estimate to ensure that errors are evaluated correctly (see
Section 2.2), but to include SIS data in the binding mass estimate to get a
better constraint on the temperature in the outer cooler region.

A Ginga LAC wide-aperture spectrum of A2163 was analyzed in Arnaud et al.\
(1992) and in EAB. Fitting the Ginga spectrum alone in the 3--30~keV band by
an isothermal model, we obtained $T_e=13.5^{+1.8}_{-1.5}$~keV with
$\chi^2=6.1/18-2$~d.o.f., fixing other parameters as
above.\footnote{Somewhat too low $\chi^2$ is due to our assumption that the
Ginga background errors in different energy channels are uncorrelated, but
since Ginga results are mostly superseded by the current ASCA measurement,
we will not concentrate on that.} Although the Ginga spectrum is consistent
with a single-temperature model, it allows the kind of non-isothermality
observed by ASCA, because only about 15\% of the overall cluster emissivity
comes from our outer shell. A simultaneous fit of the SIS, GIS and Ginga
data gives the temperatures 13.3~keV, 13.3~keV and 3.8~keV for the three
model regions, with $\chi^2=81.3/108-9$~d.o.f., in which 44.7 is due to the
SIS, 27.0 due to the GIS and 9.6 due to Ginga. Comparing these numbers to
the $\chi^2$ values obtained in separate fits for each instrument, we
conclude that there is a reasonable agreement between all instruments. We
note, however, that the overall GIS best-fit temperature, 11.5~keV, is
somewhat lower than the Ginga temperature (although they are consistent at
the 95\% level), which may have an effect on the mass estimate (for which we
used the ASCA and Ginga data jointly) comparable with its statistical error.

Finally, we have undertaken a less model-dependent analysis of the A2163
two-dimensional temperature distribution, using the actual Rosat image as a
model brightness distribution, and the result suggests that the projected
temperatures are lower in all four directions off the center (our paper in
preparation). Thus the effect we report in this paper seems to be real,
unless there remains some gross misunderstanding of the telescope
systematics. Unfortunately, due to the poor statistics far from the cluster
center, it is difficult to test this measurement with Rosat, even though the
temperatures are getting close to its 2~keV upper energy limit. Analysis of
the archival Rosat data in a ring from 3.5--7.5\am\ gives
$T_e=15^{+\infty}_{-10}$~keV and in a ring from 7.5--11.5\am\
$T_e=4^{+32}_{-2}$~keV. Thus these values are consistent with the ASCA
results but with much larger uncertainties.

\section{HUBBLE CONSTANT}

Wilbanks et al.\ (1994) detected a Sunyaev-Zel'dovich decrement toward
A2163. We have used this measurement to estimate the Hubble constant,
following the method proposed by Gunn (1978), Silk and White (1978) and
Cavaliere, Danese and De Zotti (1979), and most recently applied to A665 by
Birkinshaw, Hughes and Arnaud (1991), to A2218 by Birkinshaw and Hughes
(1994), and to Cl0016+16 by Yamashita (1994). Wilbanks et al.\ presented
their measurement in terms of the peak $y$ value for their assumed
isothermal symmetric gas model, $y_0=3.78\pm0.62\,\times 10^{-4}$ (1$\sigma$
error; a component of the error arising from uncertainties of the gas model
parameters is excluded). In the absence of the original radio scans, we
simulated them using the gas model of Wilbanks et al., and recalculated the
normalization of the SZ profile for our temperatures and the EAB's $a_x$,
$\beta$ and $\rho_{g0}$, crudely corrected for the non-isothermality (see
Section 2.1). The resulting value for the Hubble constant has a best fit of
63\kmsmpc\ and a 68\% confidence interval of 42--110\kmsmpc. The confidence
interval has been calculated by Monte-Carlo simulation, and is dominated by
the uncertainty of the currently available SZ measurement. Under the
assumption of an isothermal gas extending to $r=13\,a_x$ and a best-fit
temperature of 14.6~keV (Ginga + Rosat PSPC), the best-fit $H_0$ value is
93\kmsmpc.

\section{BINDING MASS}

Hydrodynamic simulations of the growth of clusters in a hierarchical
universe (e.g., Evrard 1990; Navarro, Frenk \& White 1995) predict
temperature profiles in equilibrium clusters generally similar to what we
have observed, a nearly isothermal inner part where most of the X-ray
emission originates, and a decline of the temperature in the outer part.
However the predicted temperature drop in the outer regions is not as steep
as we have detected. In Coma, up to date the only other cluster with the
temperature measured out to the comparable off-center distance (Hughes
1991), the temperature profile is rather similar to that in A2163.

We will now use our temperatures, supplemented by the Ginga spectrum and the
Rosat image, to try to constrain the cluster binding mass distribution,
formally applying the assumptions of hydrostatic equilibrium and spherical
symmetry, following the formalism of Hughes (1989) and Henry, Briel \&
Nulsen (1993). Although below we show that for several reasons, hydrostatic
equilibrium is an unlikely condition in the outer parts of this cluster, it
is useful to obtain the mass estimate under this assumption, for comparison
with previous estimates and to get an idea about the possible errors it can
introduce.

The gas density is assumed to have the profile
$\rho_g=\rho_{g0}\,(1+r^2/a_x^2)^{-3\beta/2}$ with $a_x$ and $\beta$ fixed
at their Rosat values, which is adequate out to large radii (EAB).  Although
non-isothermality would modify this profile derived in the assumption of the
constant temperature, this would have a minor effect on the conclusions. For
the binding mass profile of the generic form
$\rho=\rho_0\,(1+r^2/a_b^2)^{-n/2}$, we tried to find the values of
$\rho_0$, $a_b$ and $n$ consistent with the data. For such distributions of
the gas and the binding mass, the temperature radial profile is determined
by four parameters, $\rho_0$, $a_b$, $n$, and the central gas temperature
$T_0$. For several $n$ from the range of interest, a continuous temperature
profile was reconstructed for each point in the ($\rho_0$, $a_b$) plane and
compared to the ASCA+Ginga dataset (modeling the spectra in each ASCA image
ring and the overall spectrum for Ginga), fitting the remaining parameter
$T_0$ to minimize the overall $\chi^2$. We then crudely compared the
obtained temperature profile with the Rosat data (analyzed in EAB), ignoring
the Rosat spectral information and using only the fact that in its
0.4--2~keV band image, there is emission out to a certain radius, which
implies that the temperature at that radius is above zero. Thus for the
radii where the model temperature dropped below zero (which means that there
is no gravitationally bound gas there), we calculated $\chi^2$ with which
the 1\am-binned Rosat radial brightness profile differs from zero. The 90\%
constraints in the ($\rho_0$, $a_b$) plane, corresponding to $\chi^2_{\rm
min}+4.61$ for ASCA+Ginga and $\chi^2=4.61$ for Rosat, are shown in Fig.~4.

Fig.~4 shows that only a very small region of the parameter space is
marginally allowed by both datasets, namely, the region with binding mass
profiles steeper than that of the gas. Among the models allowed by
ASCA+Ginga, the Rosat data allows only those with $n\gax2.1$ and $a_b$
smaller than or equal to the gas scale length. An additional restriction is
that the total density cannot be below the gas density at the radii where
the gas is actually observed. For X-ray-measured quantities, $\rho/\rho_g
\propto h^{3/2}$, and this condition excludes the models with $n\gax2.7$ and
$n\gax3$ (among those otherwise allowed by all data) for $h=0.5$ and
$h=0.8$, respectively --- and if any amount of dark matter is to exist at
all in the cluster outskirts, this restriction should be stronger. Mass
profiles with the same slope as that observed for the gas, that is,
$n=3\beta$, which are favored by hydrodynamic
simulations\footnote{Simulations by e.g., Evrard (1990) and Navarro, Frenk
and White (1995) predict the slope of the dark matter profile only slightly
steeper than that of the gas, for the regions outside cores in equilibrium
clusters.}, are excluded at a $>$99\% confidence for all $a_b$. A
temperature profile from the allowed region for $n=2.4$ is shown in Fig.~2.
The binding density and the enclosed mass profiles which correspond to this
model are shown in Fig.~5 (for $h=0.5$), together with the gas mass from
EAB. For the allowed models, the enclosed binding mass depends only mildly
on $n$. Inside $r=0.5\,h^{-1}$~Mpc the mass is
$4.3\pm0.5\,\times10^{14}\,h^{-1}$\msun\ (of which $0.074\,h^{-3/2}$ is
gas), and inside $r=1.5\,h^{-1}$~Mpc it is
$1.07\pm0.13\,\times10^{15}\,h^{-1}$\msun\ (in which the gas fraction is
$0.14\,h^{-3/2}$, while the fraction of the gas density in the total mass
density at this radius is as high as $0.23\,h^{-3/2}$). The quoted 90\%
errors correspond to the full range of mass for the models that satisfy all
restrictions. The mass inside $r=0.75\,h^{-1}$~Mpc is slightly lower but
marginally consistent with the EAB's value, $7.3\pm0.8\,\times
10^{14}\,h^{-1}$\msun, the difference arising from the slightly different
overall temperatures used. At the larger radii, our allowed masses are
considerably lower (by a factor of 1.5 at $r=15a_x$ for the enclosed mass)
than the EAB's estimate obtained assuming isothermality. However, below
(Section 5.2) we will show that in the outer part, the gas hydrostatic
equilibrium assumption is invalid and we are likely to underestimate the
mass at the larger radius.

\section{DISCUSSION}

\subsection{The overall temperature}

The high value of the cluster overall temperature obtained by Ginga and ASCA
is in line with the cluster's high X-ray luminosity and its velocity
dispersion (Arnaud et al.\ 1992; Soucail et al.\ 1995; Lubin \& Bahcall
1993; however, its velocity dispersion may be overestimated because of the
substructure). This temperature is not affected by the presence of the
diffuse radio halo nor the existence of a possible AGN in the cluster.

Herbig and Birkinshaw (1995) discovered a radio halo in this
cluster, the most luminous and extensive halo of those detected, and
estimated that as much as 10\% of the cluster's X-ray emission in the
0.5--4.5~keV band may be produced by the inverse Compton scattering of
cosmic microwave background photons by the relativistic electrons of the
halo. If the relativistic electrons have the power-law energy distribution
over a range of energies, and the halo has the typical steep radio spectrum
(so that its X-ray photon index is $\alpha_{\rm xray}=\alpha_{\rm radio}
\approx 2.2$, e.g., Jaffe 1977; review in Sarazin 1988), an inverse Compton
contribution to the cluster X-ray spectrum is expected to be between 4--7\%
over the energy band 5--20~keV, therefore, the temperature measurement
should be essentially unaffected. When we limit the flux of the power law
component to less than 7\% of the total flux in our energy band and let its
index vary, the best fit temperature in the inner shell (using GIS) is
$10.6^{+1.8}_{-1.9}$~keV at 68\% confidence, consistent with the model
without a power law component. If we fix the index at a value of 2.2 the
upper limit on the power-law contribution to the total flux in our band is
3.5\% (90\% confidence).

The possibility of a significant AGN contribution to the cluster X-ray
spectrum was considered by Arnaud et al.\ (1992), who included a non-thermal
component with the AGN spectrum in their fit to the Ginga data in the
2--20~keV energy band, and concluded that such component should be
negligible. From a spectral fitting point of view the possibility of an
additional AGN component is similar to that of the diffuse component
considered above and results in similar upper limits. Furthermore, a
detection of the $H$-like iron $K\alpha$ line in the ASCA SIS cluster
spectrum (Paper I) confirms that the gas is indeed very hot --- the $H$-like
to $He$-like iron line ratio limits the temperature of the gas emitting iron
lines to at least 12~keV, assuming ionization equilibrium.

\subsection{The mass distribution}

The analysis presented in Section 4, made under the hydrostatic equilibrium
assumption, may be interpreted as showing that the binding mass distribution
is well restricted out to the large off-center distances, and most
interestingly, one can rule out a the distribution of the dark matter
similar to that of the baryonic gas. However, it can be seen from Fig.~2,
which also shows one of the excluded temperature profiles from the family of
models in which mass has the same slope as gas, that it is excluded mainly
because the observed temperature is too low in our outer radial bin, while
Rosat detects emission at a greater radius. Although, as is shown in Section
4, there is still room for some of the more exotic total mass distributions
(which assume even more exotic dark matter profiles), we note that there are
at least a few natural reasons for the observed temperature to be lower than
that predicted by an equilibrium model. An estimate (e.g., Fabian et al.\
1986) shows that the observed steep temperature drop between the middle and
the outer radial shells implies that they are convectionally unstable for
any polytropic index $\gamma<1.8$. Even if convection is slow and local
pressure equilibrium is always achieved, the measured temperature in the
outer cluster part would give an underestimate of the total pressure, since
bulk motions of the gas would account for some fraction of it.

One also notes that although A2163 does not exhibit the prominent
substructure that is observed in many other clusters, it may still have
recently undergone (or be undergoing) a merger, as suggested by Soucail et
al.\ (1995) from the galaxy velocities, by EAB from the morphology of the
Rosat image, and in our Paper~I from the small-scale non-isothermalities
near the cluster center. Simulations suggest (e.g., Navarro, Frenk \& White
1995) that a high value of $\beta_T \equiv \mu m_p \sigma_{\rm gal}^2/kT$,
which is 1.3--1.5 for this cluster (from the data in the references above),
also is a property of non-relaxed clusters. If the merger is still going on,
the gas outside the cluster hot central region may not have settled down and
is continuing to inflow with the bulk velocities as high as 1000--2000\kms,
as some merger simulations predict (see, e.g., Fig.~3{\em e} in Schindler \&
\muller\ 1993). If such flows exist, the gas kinetic energy can be as high
as $\mu m_p \sigma^2_{\rm gas}/2 \sim 3-5$~keV per particle, comparable with
its thermal energy, making the hydrostatic equilibrium assumption invalid in
the outer cooler region. The measured temperature would then again lead to
an underestimate of the total pressure at those distances.

Another possible result of a recent merger is absence of electron-ion
equipartition. At the distance of $10a_x$ in A2163, the electron density is
about $10^{-4}\,h^{1/2}$\cmcube\ (EAB), and the timescale for reaching
electron-ion temperature equality via collisions is
$1-2\,\times10^9\,h^{-1/2}$~yr (Spitzer 1956), comparable to the merger
timescale. Thus for the gas heated by shock waves during a merger, the
measured electron temperature would give an underestimate of the gas
pressure at those distances shortly after the merger. Note that for electron
densities and temperatures in the range of interest, the characteristic time
of change of the ion fractions for highly ionized iron is about an order of
magnitude shorter than the above timescale (e.g., Mewe \& Gronenschild
1981), therefore the emission in the iron spectral lines should be
consistent (at least with the present experiment accuracy) with the electron
temperature, and will not indicate that electrons and ions are out of
equipartition. The SZ effect (Sunyaev \& Zel'dovich 1972) is proportional to
the electron temperature and will not show it either. Fig.~2 shows that an
underestimate of the thermodynamic temperature in our outer shell by a
factor of $\gax 2$ could produce the observed discrepancy with the model in
which dark mass has the distribution similar to that of gas (and also result
in the conclusion about convectional instability). The mass profile which
corresponds to this excluded model (dotted lines in Fig.~5) illustrates that
such an underestimate of the outer temperature would have little effect on
our total mass estimate at the smaller radius, while the mass within
$r=10a_x$ may be significantly underestimated.

The discovery of the strongest radio halo in this hottest cluster may have
another implication relevant to our discussion. All radio halos are found in
hotter clusters, and it is not ruled out that the energy spectrum of the
halo's relativistic electrons extends to low enough energies to efficiently
heat the intracluster gas (Lea \& Holman 1978; Vestrand 1982; Sarazin 1988
has a discussion), with the heating rate that may even exceed the X-ray
radiation cooling. If such heating is indeed present, the gas in the inner
region may have insufficient time to equalize its increasing pressure with
the outer shells. Such regime would require a net energy supply over the
region occupied by the halo ($1.2\,h^{-1}$~Mpc, Herbig \& Birkinshaw) about
an order of magnitude higher than the cluster X-ray luminosity. Since there
is no apparent ways to test it observationally, we only mention that it may
lead to an overestimate of the binding mass in the cluster inner part. A
weak extended magnetic field whose presence is indicated by the radio halo,
should have an energy density negligible comparing to the gas pressure and
therefore will not affect the mass determination on the large linear scale.

To conclude, although we did not explore the effects of the cluster
asymmetry and possibly more complex temperature structure in detail (such a
study in now under way), the observed discrepancy between the data and the
simple model where dark matter and gas have similar distributions, most
naturally indicates that some of the gas equilibrium assumptions do not hold
in the outer part of this cluster.

\section{SUMMARY}

We have performed a spatially-resolved temperature measurement of Abell
2163, the hottest of known clusters, and found that it is cooler beyond
$1\,h^{-1}$~Mpc from the center, similar to Coma and to the hydrodynamic
simulation predictions. Combining our temperature measurements with the
Rosat imaging and the Ginga wide-aperture spectrum, we were able to severely
restrict the possible binding mass distributions of the form
$\rho=\rho_0\,(1+r^2/a_b^2)^{-n/2}$, if not exclude them all, formally
applying the assumption that the gas is in hydrostatic equilibrium at the
measured temperatures. However, this assumption is likely to be inadequate
in this non-relaxed cluster at large off-center distances, where, for
example, convection should exist because of the steep temperature drop, or
bulk gas motions are expected during a merger, or the electron-ion
temperature equipartition timescale becomes comparable to the merger
timescale. If true, this would lead to a significant underestimate of the
gravitating mass density in the cluster outskirts, leading to large apparent
values of the baryonic density. We have also estimated $H_0$ using the ASCA
electron temperatures and the previously reported Rosat density profile and
the Sunyaev-Zel'dovich decrement. The obtained value is poorly constrained,
however, due to the large uncertainty of the available SZ measurement.

\acknowledgments

We are grateful to all members of the ASCA team for the continuous support.
We thank A.~Fabian for interesting discussion, and the referee for many
useful suggestions. M.M. would like to thank ISAS for its support and
hospitality during this work.

\appendix
\section{MODELING THE TELESCOPE RESPONSE WITH WIDE PSF}

As we have mentioned in Section 2.2, the wide energy- and position-dependent
ASCA mirror PSF requires that the spectra from different image regions be
fit simultaneously. To do this within a reasonable computing time, we have
used the following simple scheme. Denoting the projected sky image of the
emission measure of the $i$-th model region (either two- or
three-dimensional) as $m_i$, its spectrum as $s_i(E)$ (where $E$ is in keV),
the total number of the model regions as $M$, the operation of summing the
flux over the $j$-th detector image region as $R_j$ (including in it the SIS
gaps and the GIS grid for clarity), the spectrum from the $j$-th detector
image region as $d_j(E')$ (where $E'$ is in channels), the total number of
the detector image regions as $N$, the linear operation of multiplying by
the mirror effective area plus PSF scattering, which converts a model
brightness distribution into that in the detector plane, as $P(E)$, and
convolution of the spectrum with the coordinate-independent detector
spectral response including the detector efficiency, as $D$, we have
\[
d_j= D\, R_j \sum_{i=1,M} P\, m_i\, s_i.
\]
The telescope response transform, $T_{ij}(E)\equiv R_j \sum_{i=1,M} P(E)\,
m_i$, which for a given energy converts the spectra of the model regions to
the fluxes in the detector image regions, \mbox{$d=D\,Ts$}, is thus an $M
\times N$ matrix that includes geometry of the source and
the integration regions, vignetting and PSF scattering. The least-squares
solution $s$ is searched by iterations. If there is no PSF scattering and no
projection, $T$ is diagonal and spectra from the different image regions can
be fit separately. If $m_i$ can be fixed or reasonably simply parametrized
(for example, a Rosat image is available or a $\beta$-model brightness
profile can be assumed), $T(E)$ can be calculated once in the beginning
(which involves time-consuming two-dimensional convolutions with the
position-dependent PSF), after which only the spectral parameters and maybe
relative normalizations of the model spectra need to be fit, each iteration
involving only the convolution with the small matrix $T$ (for reasonable $M$
and $N$). In the fit of our three temperatures assuming a fixed
$\beta$-model emission measure profile, $T$ was a $3\times 3$ matrix for
each energy. To fit the density profile parameters in addition to the
temperatures, we have calculated the transform for the 16 1\am-wide radial
model shells (that is, $T$ was a $16\times 3$ matrix), assuming uniform
emission measure throughout the cluster, and only recalculated the
normalization for each shell in each iteration according to the values of
$a_x$ and $\beta$, taking advantage of linearity of all the involved
operations.

The telescope PSF is poorly known at the moment and an estimate of its
systematic error has been included in the analysis. We assumed a 5\%
$1\sigma$ error (Takahashi et al.\ 1995; see Section 2.2) of the
non-diagonal elements of the matrix $T$. In Monte-Carlo calculation of the
confidence intervals of the model temperatures (Section 2), a 5\% random
error has been added directly to the elements of $T$. For calculation of all
the $\chi^2_{\rm min}$ values and confidence contours presented in the
paper, an approximate equivalent of this error, $\sigma_j$, has been added
in quadrature to the statistical error of the data value $d_j$:
\[
\sigma_j^2=\sum_{i\neq j} \left( 0.05\, d_i\, \frac{T_{ij}}{T_{ii}} \right)^2
\]
(which is a slight overestimate due to the neglect of non-diagonal elements
of $T$). Here $T$ is a matrix calculated for the adequate emission measure
distribution and the model regions which are the same as the data regions.
If the data rings are narrower than, say, 3--4\am, an error should be added
to the diagonal elements of $T$ as well.

\clearpage

\section*{Figure Captions}

\vspace{.5cm}

Fig.~1.---Spectra of the three image regions (in detector counts) and the
best-fit model (see Fig.~2). Panels correspond to the four SIS chips and the
two GISs whose data were fit simultaneoulsy. Upper, medium and lower spectra
correspond to central, medium and outer image rings, respectively (values
for the central and medium rings are multiplied by factors 10 and 2,
respectively, for clarity). Plotted errors are $1\sigma$, including errors
of the background but not including the PSF systematics.

\vspace{.5cm}

Fig.~2.---Deprojected temperatures in three spherical regions of the
cluster, obtained with GIS+SIS, are shown as crosses. Vertical bars are
single-parameter 90\% intervals which include all known systematic errors.
Crosses are centered at the $EM$-weighted radius for each bin. For the outer
region, spectra were collected from the ring with the outer radius smaller
than that used for the model, which is shown by the dashed line. The arrow
corresponds to (at least) a 90\% lower limit on the temperature at about
$13a_x$, which arises from the Rosat detection of the cluster emission at
that radius. The smooth solid line shows one of the allowed temperature
profiles (that for $n=2.4$ and $a_b=0.5\,a_x$) under the hydrostatic
equilibrium assumption. A temperature profile for the mass model with the
same $a_b$ as above but with the slope similar to that of gas,
$n=3\beta=1.9$, excluded at a $>99$\% confidence, is shown as a dotted line.

\vspace{.5cm}

Fig.~3.---Confidence contours for the density profile parameters $a_x$ and
$\beta$, obtained letting temperatures in the three model shells vary freely
and fitting the data from the two GISs. The contours correspond to 68\%
($\chi^2_{\rm min}+2.3$) and 90\% ($\chi^2_{\rm min}+4.61$) joint
two-parameter confidence. A cross marks the Rosat best-fit values for the
isothermal fit ($a_x=1.2'$, $\beta=0.62$). Our best-fit values are
$a_x=1.4'$, $\beta=0.66$, and the temperatures 12.2~keV, 12.7~keV and
3.7~keV, with $\chi^2_{\rm min}$=21.8/30--6~d.o.f.

\vspace{.5cm}

Fig.~4.---Constraints on parameters of the binding mass distribution, for
three fixed values of $n$. Regions inside the solid contours are allowed by
the ASCA+Ginga data at the 90\% confidence, while the Rosat imaging data,
for the same value of the remaining parameter $T_0$, exclude the region
above the dashed curve at the 90\% confidence at least. The regions
marginally allowed by both datasets are shaded. Models with $n\gax3$ or
$n\gax2.7$ are excluded for $h=0.8$ or $h=0.5$, respectively, because the
total density becomes lower than the observed gas density at some radii in
those models.

\vspace{.5cm}

Fig.~5.---Radial profiles of the gas (from EAB) and the total mass, for
$h=0.5$. Descending lines show density and ascending lines show enclosed
mass. Solid line corresponds to the total mass model with $n=2.4$ and
$a_b=0.5\,a_x$, allowed by the constraints in Fig.~4 under the hydrostatic
equilibrium assumption. Error bars at $r=3.3a_x$ and $9.8a_x$ correspond to
the 90\% confidence intervals on the total mass, calculated for the models
complying with all constraints. However, hydrostatic equilibrium is likely
to be broken in the cluster outer part; see Section 5.2. Dotted line
corresponds to one of the excluded models in which the mass profile has the
same slope as that of gas.

\clearpage
\pagestyle{empty}

{\bf Fig. 1}
\vspace{3cm}

\centerline{\epsfxsize=16cm
\includegraphics{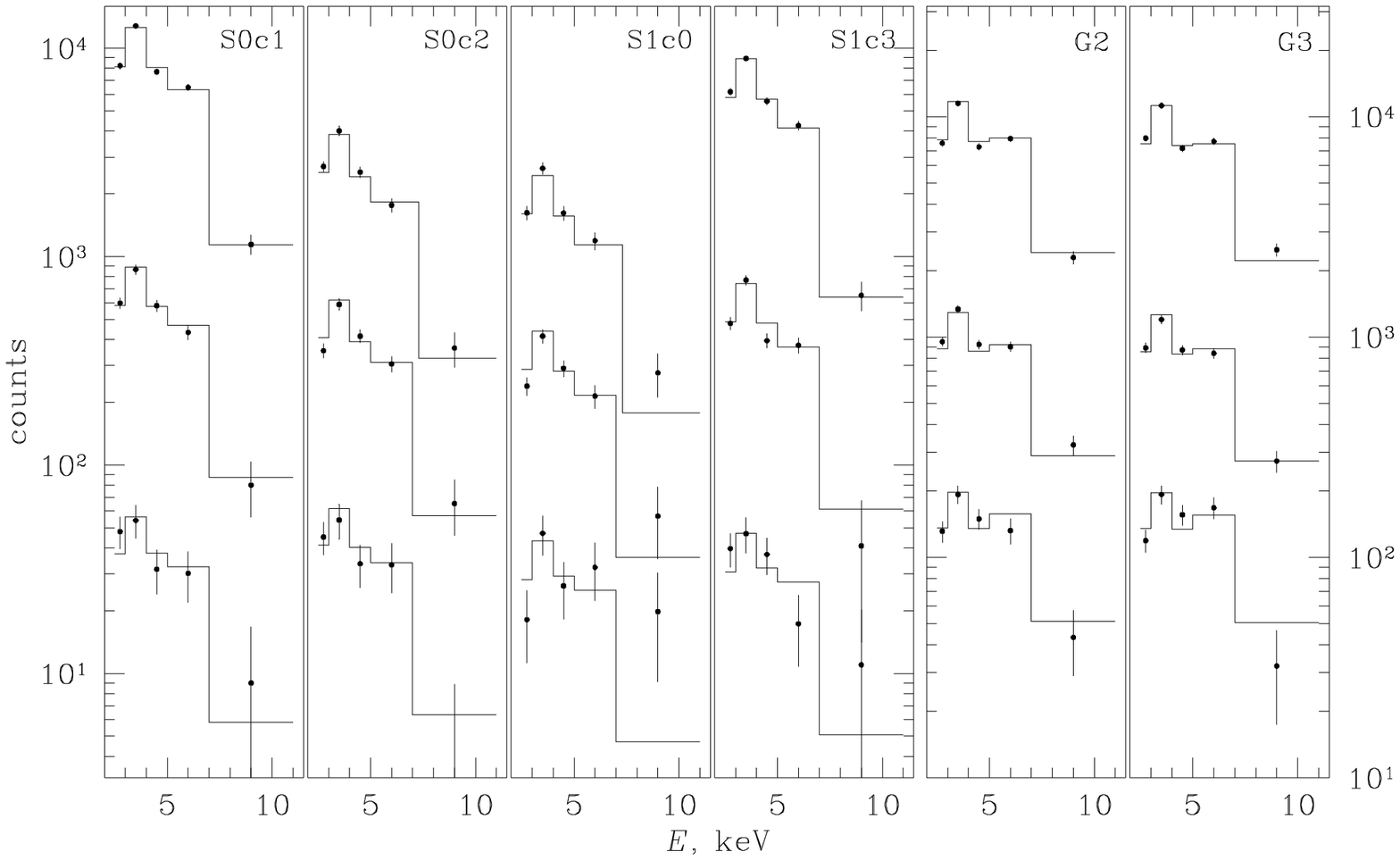}}

\clearpage
\pagestyle{empty}

{\bf Fig. 2}
\vspace{3cm}

\centerline{\epsfxsize=16cm \epsffile{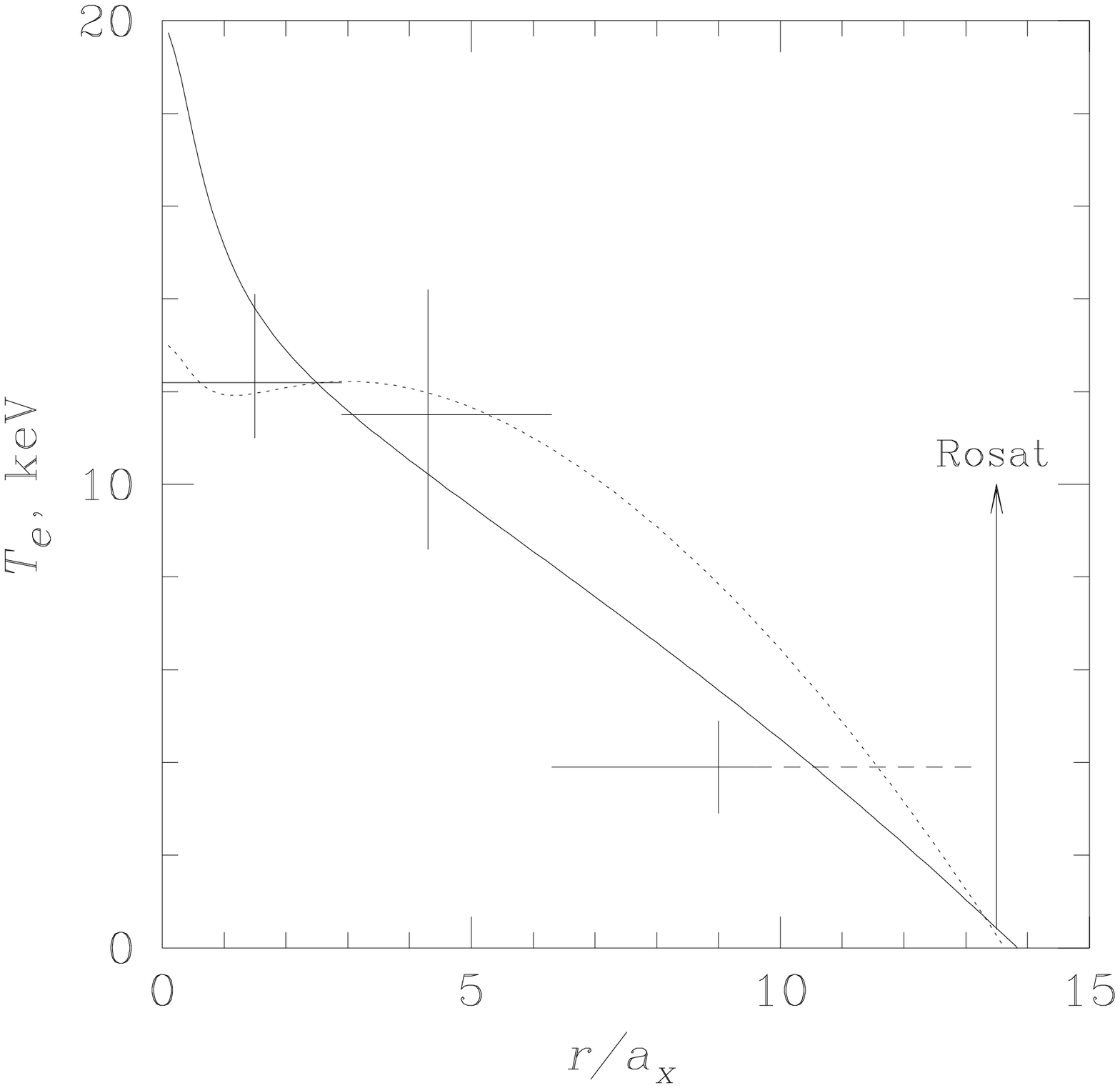}}

\clearpage

{\bf Fig. 3}
\vspace{3cm}

\centerline{\epsfxsize=16cm \epsffile{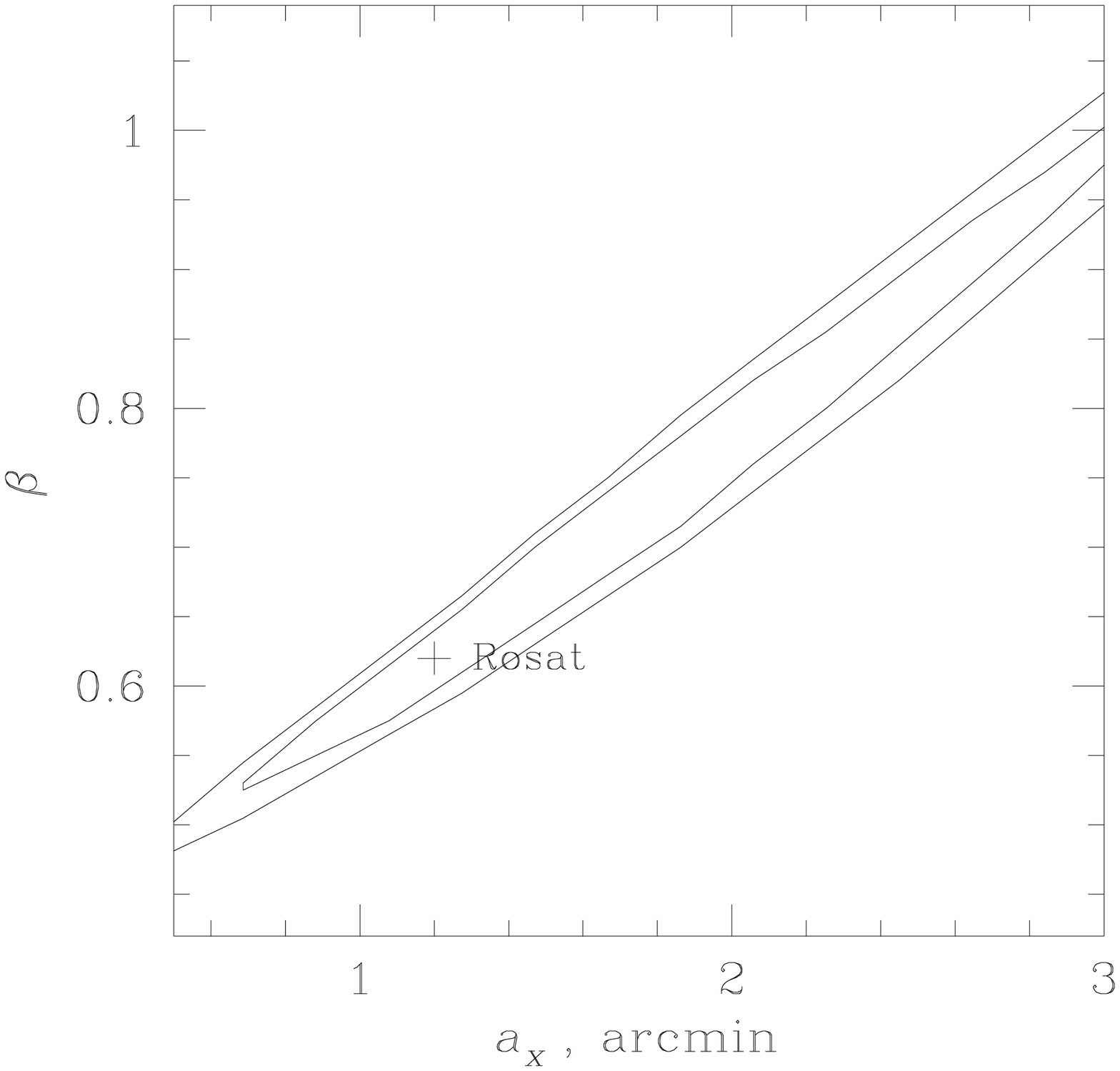}}

\clearpage

{\bf Fig. 4}
\vspace{3cm}

\centerline{\epsfxsize=16cm \epsffile{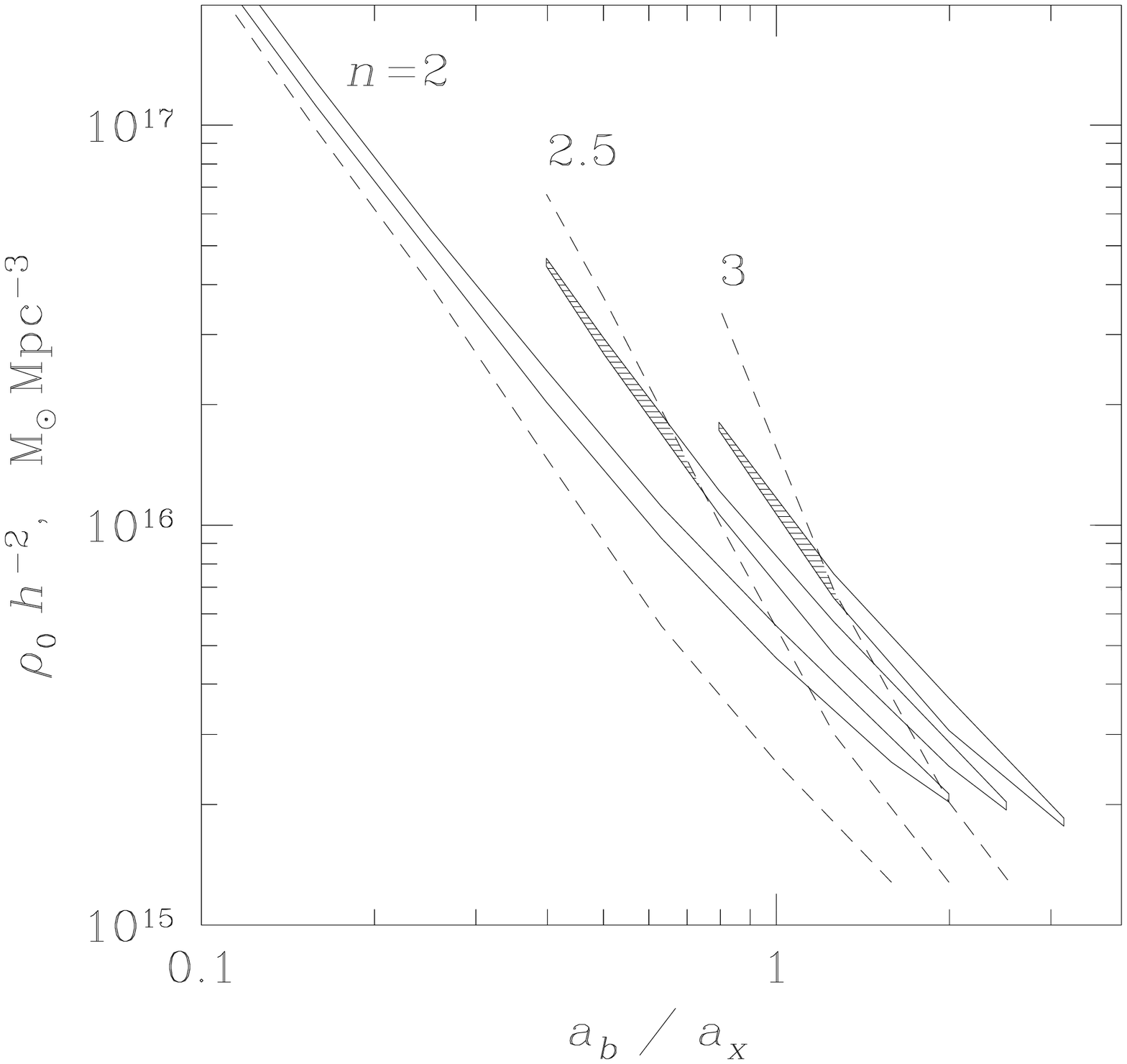}}

\clearpage

{\bf Fig. 5}
\vspace{3cm}

\centerline{\epsfxsize=16cm \epsffile{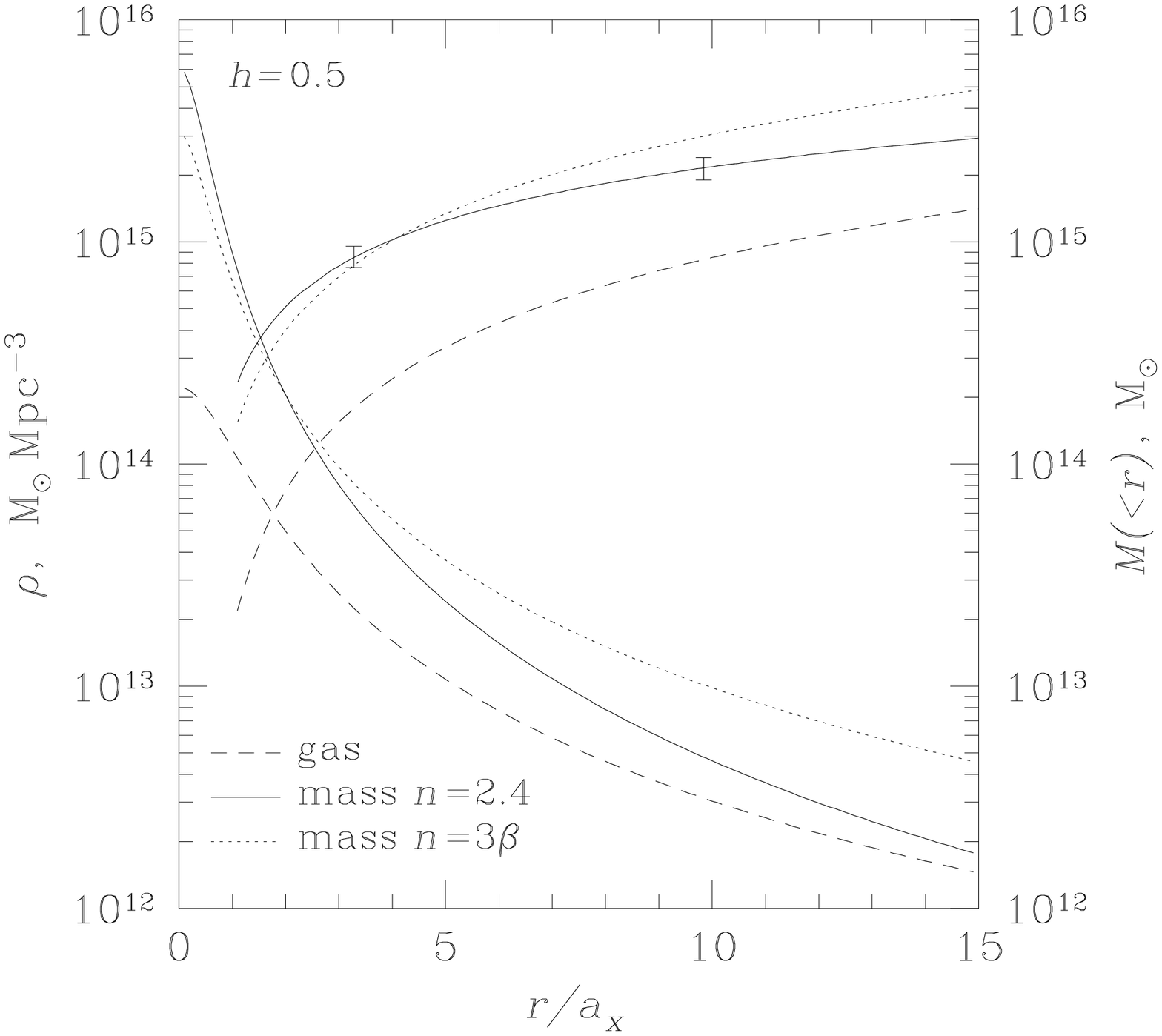}}


\clearpage

\begin{references}

\reference Arnaud, M., Hughes, J.P., Forman, W., Jones, C.,
Lachieze-Rey, M., Yamashita, K., \& Hatsukade, I., 1992, ApJ, 390, 345

\reference Arnaud, M., Elbaz, D., \bohringer, H., Soucail, G., \& Mathez, G.
1994, in New Horizon of X-Ray Astronomy, ed.\ F.  Makino \& T. Ohashi
(Tokyo: Universal Academy Press), 537

\reference Birkinshaw, M., Hughes, J.P., \& Arnaud, K.A. 1991, ApJ, 379, 466

\reference Birkinshaw, M. \& Hughes, J.P. 1994, 420, 33

\reference Briel, U.G., \& Henry, P. 1994, Nature, 372, 439

\reference Cavaliere, A., Danese, L., \& De Zotti, G. 1979, A\&A, 75, 322

\reference Daines, S., Jones, C., Forman, W., \& Tyson, A. 1995, ApJ,
submitted

\reference Elbaz, D., Arnaud, M., \& \bohringer, H. 1995, A\&A, 293, 337 (EAB)

\reference Evrard, A.E. 1990, ApJ, 363, 349

\reference Fabian, A.C., Thomas, P.A., Fall, S.M., \& White, R.E. 1986,
MNRAS, 221, 1049

\reference Fabricant, D., Rybicki, G., \& Gorenstein, P. 1984, ApJ, 286, 186

\reference Gunn, J.E. 1978, in Observational Cosmology, ed.\ A. Maeder et
al.\ (Sauverny: Geneva Observatory), 1

\reference Henry, J.P., Briel, U.G., \& Nulsen, P.E.J. 1993, A\&A, 271, 413

\reference Herbig, T, \& Birkinshaw, M. 1995, BAAS, 26, 1403

\reference Hughes, J. 1989, ApJ, 337, 21

\reference Hughes, J. 1991, personal comm., presented in Clusters and
Superclusters of Galaxies, ed. A.C. Fabian (Dordrecht: Kluwer), 54

\reference Ikebe, Y. 1995, PhD Thesis, Tokyo University

\reference Jaffe, W.J. 1977, ApJ, 212, 1

\reference Jones, C., \& Forman, W. 1984, ApJ, 276, 38

\reference Kubo, H. 1994, personal comm.

\reference Lea, S.M., \& Holman, G.D. 1978, ApJ, 222, 29

\reference Loeb, A., \& Mao, S. 1994, ApJ, 435, L109

\reference Lubin, L.M., \& Bahcall, N.A. 1993, ApJ, 415, L17

\reference Markevitch, M., Yamashita, K., Furuzawa, A., \& Tawara, Y. 1994,
ApJ, 436, L71 (Paper I)

\reference Mewe, R., \& Gronenschild, E.H.B.M. 1981, A\&A Suppl., 45, 11

\reference Mitchell, R., Culhane, J., Davison, P. and Ives, J. 1976, MNRAS,
189, 329

\reference Mushotzky, R.F., Serlemitsos, P.J., Smith, B.W., Boldt, E.A., \&
Holt, S.S. 1978, ApJ, 225, 21

\reference Navarro, J.F., Frenk, C.S., \& White, S.D.M. 1995, MNRAS, in press

\reference Raymond, J.C., \& Smith, B.W. 1977, ApJ Suppl., 35, 419

\reference Sarazin, C.L. 1988, X-ray Emission from Clusters of Galaxies
(Cambridge: Cambridge University Press)

\reference Schindler, S., \& M\"{u}ller, E. 1993, A\&A, 272, 137

\reference Serlemitsos, P., et al. 1995, in preparation

\reference Silk, J.I., and White, S.D.M. 1978, ApJ, 226, L103

\reference Soucail, G., Arnaud, M., \& Mathez, G. 1995, in preparation

\reference Spitzer, L.Jr. 1956, Physics of Fully Ionized Gases (New York:
Interscience)

\reference Sunyaev, R.A., \& Zel'dovich, Ya.B. 1972, Comments Astrophys.
Space Phys., 4, 173

\reference Takahashi, T., Markevitch, M., Fukazawa, Y., Ikebe, Y., Ishisaki,
Y., Kikuchi, K., Makishima, k., \& Tawara, Y. 1995, ASCA Newsletter, no. 3
(NASA/GSFC)

\reference Wilbanks, T.M., Ade, P.A.R., Fischer, M.L., Holzapfel, W.L., \&
Lange, A.E. 1994, ApJ, 427, L72

\reference Yamashita, K. 1994, in New Horizon of X-Ray Astronomy, ed.\ F.
Makino \& T. Ohashi (Tokyo: Universal Academy Press), 279

\end{references}
\end{document}